\documentclass{article}
\usepackage{amsmath,amsthm}
\usepackage{graphicx,epstopdf,epsfig,multirow,epic,bm}
\usepackage{subfigure}
\usepackage{amsfonts}

\normalsize \rm
\begin{document}

\begin{center}
{\Large  \textbf { Dense networks with scale-free feature }}\\[12pt]
{\large Fei Ma$^{a,}$\footnote{~The author's E-mail: mafei123987@163.com. },\quad Xiaomin Wang$^{a,}$\footnote{~The author's E-mail: wmxwm0616@163.com. },\quad  Ping Wang$^{b,c,d,}$\footnote{~The corresponding author's E-mail: pwang@pku.edu.cn.} and \quad Xudong Luo$^{e,}$\footnote{~The author's E-mail: luoxudong117@163.com. }}\\[6pt]
{\footnotesize $^{a}$ School of Electronics Engineering and Computer Science, Peking University, Beijing 100871, China\\
$^{b}$ National Engineering Research Center for Software Engineering, Peking University, Beijing, China\\
$^{c}$ School of Software and Microelectronics, Peking University, Beijing  102600, China\\
$^{d}$ Key Laboratory of High Confidence Software Technologies (PKU), Ministry of Education, Beijing, China\\
$^{e}$ School of Mathematics and Statistics, Lanzhou University, 730000 Lanzhou, China}\\[12pt]
\end{center}

\begin{quote}
\textbf{Abstract:}While previous works have shown that an overwhelming number of scale-free networks are sparse, there still exist some real-world networks including social networks, urban networks, information networks, which are by observation dense. In this paper, we propose a novel framework for generating scale-free graphs with dense feature using two simple yet helpful operations, first-order subdivision and Line-operation. From the theoretical point of view, our method can be used not only to produce desired scale-free graphs with density feature, i.e. power-law exponent $\gamma$ falling into the interval $1<\gamma\leq2$,  but also to establish many other unexpected networked models, for instance, power-law models having large diameter. In addition, the networked models generated upon our framework show especially assortative structure. That is, their own Pearson correlation coefficients are able to achieve the theoretical upper bound. Last but not the least, we find the sizes of community in the proposed models to follow power-law in form with respect to  modularity maximization.\\


\end{quote}

\vskip 1cm

\section{Introduction}

Complex networks, usually interpreting diverse complex systems around us, have attracted more attention in the past. Studied example networks include the Internet and the World Wide Web \cite{Albert-1999}, scientific citation and collaboration \cite{Pierre-2019}, sexual contact network \cite{David-2017}, metabolic network \cite{Juliane-2014}, and predator-prey chain \cite{Ian-2015}, to name just a few. It is a convention for one to denote a networked model by a graph $\mathcal{G}(\mathcal{V},\mathcal{E})$ which, in the simplest form, is a set of vertices in $\mathcal{V}$, representing individual members of model, joined together in pairs by edges in $\mathcal{E}$, indicating relationships between members. With such a representation, many intriguing properties planted in the topological structure of networks have been unveiled, for instance, small-world property \cite{Watts-1998}, power-law degree distribution (i.e., the so-called scale-free feature) \cite{Albert-1999-1}, community structure \cite{Newman-2006}, self-similarity \cite{Song-2005}, assortative mixing \cite{Newman-2002}. To better understand the generation principles which control or produce the emergence of characters mentioned above, a wide range of technical methods have been developed and then used to establish a large variety of theoretical models. For example, the well-known WS-model was proposed by Watts \emph{et al} \cite{Watts-1998} to try to explain small-world phenomena in various real-world networks using two measures, diameter (or average path length) and clustering coefficient. For probing mechanisms governing the scale-free feature, a great deal of models have been constructed based on various thoughts and however there seems to be no a complete consensus in current science community. Among which, the most prominent of widely studied networked models is the BA model built by Barabasi \emph{et al} \cite{Albert-1999-1} using two rules, \emph{growth and preferential attachment}, where newly added vertex tends to connect with higher probability to highly connected vertices. Throughout this paper, all graphs (models) addressed are simple and the terms \emph{graph} and \emph{network} are used indistinctly.

While a large number of networked models have been generated for modeling real-life networks, the most attractive of them are scale-free networked models. As previously, the extensive study of these such models triggers the blossom of scale-free graphs study. In 2003, through measuring the diameter $D$ or average path length $APL$ on scale-free networks with vertex number $|\mathcal{V}|$ and degree distribution $P(k)\sim k^{-\gamma}, \gamma\in(2,3]$, Cohen \emph{et al} in \cite{R-C-2003} proved using analytical arguments that these networks are small, i.e., $D\sim\ln |\mathcal{V}|$ when $\gamma\in(2,3)$, and even ultrasmall, i.e. $D\sim\ln\ln |\mathcal{V}|$ when $\gamma=3$. In fact, there are scale-free graphs with power-law exponent $\gamma=3$ whose diameters is much larger than mentioned above, see \cite{Ma-2019} for more detail. As will be shown shortly, the large diameter phenomena can also be found on the following scale-free graphs with density structure, which is the main topic of this paper.

For a theoretical networked model $\mathcal{G}(\mathcal{V},\mathcal{E})$ with an infinity of vertices, it is easy to show by definition of average degree $\langle k\rangle=2|\mathcal{E}|/|\mathcal{V}|$ whether that graph is sparse or not. Sparse models show $\langle k\rangle \rightarrow \alpha$ in the limit of large graph size and the dense ones are of $\langle k\rangle \gg \alpha$ where $\alpha$ is a constant. Obviously, all scale-free networks with $\gamma$ more than $2$ are sparse. In 2011, based on extreme value arguments, Genio \emph{et al} showed both numerically and analytically that the probability for finding a scale-free network with a given $\gamma (\in[0,2])$ is $0$ \cite{C-I-D-G-2011}. Therefore, they demonstrated that all scale-free networks have sparsity structure. As we will show later, scale-free graphs with exponent $\gamma=2$ may be easily constructed using a novel framework proposed in this paper. In addition, other scale-free graphs whose exponents $\gamma$ are in the interval from $1$ to $2$ are able to be completely generated in terms of our framework.

The rest of this paper can be organized by the following several sections. In Section II we lay out the principled framework for producing the desirable networked models of great interest. And then, the goal of Section III is to in-detail discuss some widely-studied topological properties including degree distribution and clustering coefficient on the models generated using our framework. Finally, we briefly describe our conclusions in Section IV.

\section{Introduction to framework for producing dense graphs }

In practice, the dense scale-free networks have been paid little attention in the whole history of scale-free network studies. However, according to both many real-world example networks and some fresh instruments in the literature for instance \cite{Leskovec-2007} and \cite{Lambiotte-2016}, this branch begins to become active. Most generally, the simplest method for densifying a sparse graph $\mathcal{G}$ with a given vertex number $|\mathcal{V}|$ is to consecutively add new edges to connect some vertex pairs which are not connected previously. While such an implementation can easily achieve a desirable dense graph $\mathcal{G}'$, some interesting structural properties rooted on initial graph $\mathcal{G}$ may be destroyed thoroughly. As an immediate example, the quantities closely associated with topological structure of graph $\mathcal{G}$ can be first damaged, for instance, degree distribution and diameter. Thus, one should attempt to directly produce some desirable graphs with both density structure and many other topological properties of scientific interest, such as, power-law degree distribution. Lambiotte \emph{et al} in \cite{Lambiotte-2016} introduced a minimal generative model, named the copying model, for densifying networks $\mathcal{G}(\mathcal{V},\mathcal{E})$ in which a new vertex attaches to a randomly selected target vertex and also to each of its neighbors with probability $p$. Based on rate equation approach and some additional assumptions, they have proven analytically that in some cases, these networked models may follow the power-law degree distribution with exponents $\gamma$, where $\gamma$ satisfies the following equation

\begin{equation}\label{MF-1}
\gamma=1+p^{-1}-p^{\gamma-2}.
\end{equation}
As pointed by the same authors in Ref.\cite{Bhat-2016}, in the dense regime, many features of the degree distribution of networked models mentioned above become anomalous. For instance, the degree distribution does not self-average but appears to slowly converge to a form that is close to, but distinct from, a log-normal in the large graph size limit.

In brief, how to effectively construct a dense graph with scale-free structure is a challenging and intriguing problem. In order to address this issue, we will design a novel framework for densifying sparse graphs. Different from those schemes listed above, our framework is not to directly produce a dense graph whose degree distribution has a power-law form but to generate a desirable graph on the basis of a sparse graph. In fact, it contains two components as follows.

\emph{Construction}---Here, we develop a theoretical framework for switching a scale-free graph with sparsity structure into a candidate graph with density structure. As mentioned above, we will introduce our framework in two stages. First, a well-studied operation $f$ from graph theory, named \emph{Line-operation}, is to transform an initial graph $\mathcal{G}(\mathcal{V},\mathcal{E})$ to the corresponding line graph, also called edge graph, denoted by $\mathcal{G'}(\mathcal{V'},\mathcal{E'})$. As an illustrative example, Fig.1 shows such an operation $f$ transforming a sparser graph in the left panel into a denser one in the right panel.

\begin{figure}
\centering
\includegraphics[height=5.5cm]{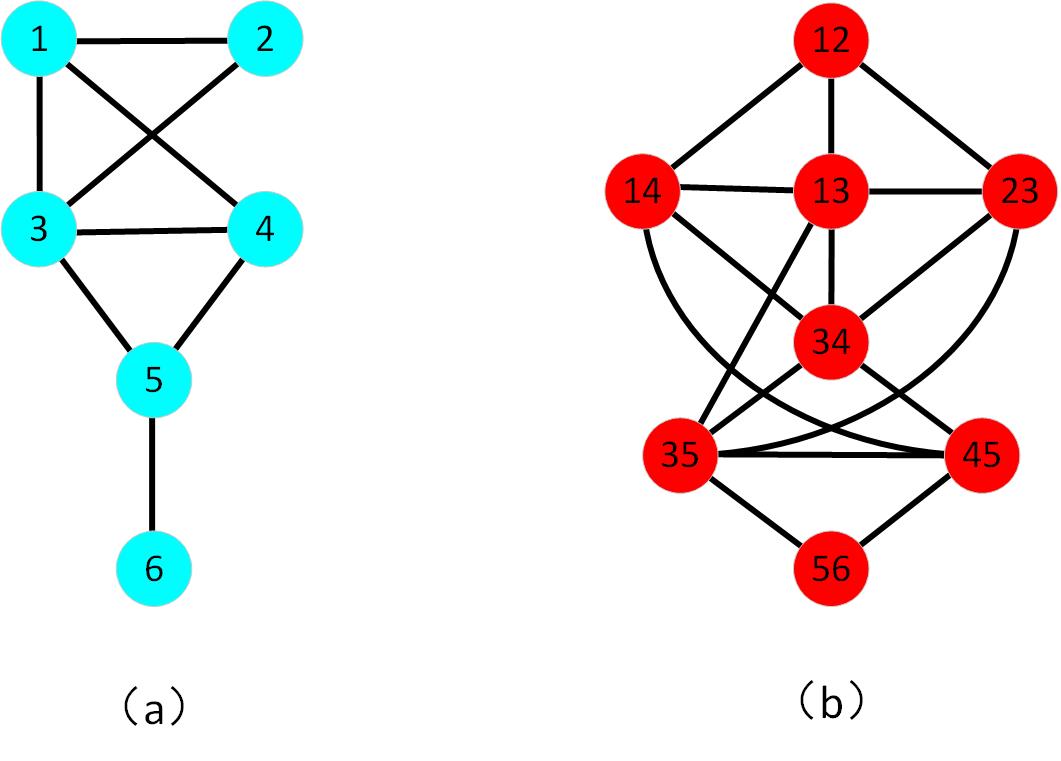}
\caption{\label{fig:epsart} The switching procedure from a sparse scale-free graph into a denser scale-free graph. For a graph $\mathcal{G}(\mathcal{V},\mathcal{E})$, its line graph based on Line-operation $f$ can be defined in the following form to be $\mathcal{G'}(\mathcal{V'},\mathcal{E'})$ whose vertex set is $\mathcal{V'}=\{v_{e}|f(e)=v_{e}, \forall e\in \mathcal{E}\}$ in which the mapping $f$ is to transform an edge to a unique vertex. Two vertices $v_{e_{i}}$ and $v_{e_{j}}$ in $\mathcal{V'}$ are adjacent if the corresponding edges $e_{i}$ and $e_{j}$ in $\mathcal{E}$ are adjacent in $\mathcal{G}(\mathcal{V},\mathcal{E})$. These such edges constitute the edge set $\mathcal{E'}$. Here, average degree $\langle k\rangle$ of graph in Fig.1(a) is equal to $4/3$ and average degree $\langle k'\rangle$ of its line graph shown in Fig.1(b) is $2$. Apparently, $\langle k'\rangle$ is greater than $\langle k\rangle$, implying that Line-operation $f$ indeed achieves the transformation from sparser graphs into denser ones.}
\end{figure}

Consider a given graph $\mathcal{G}(\mathcal{V},\mathcal{E})$ built by the probability generating function
\begin{equation}\label{MF-3}
G(x)=\sum_{k=0}^{\infty}p_{k}x^{k}
\end{equation}
here $p_{k}$ is the fraction of vertices with degree $k$, its average degree is able to be expressed as $\langle k\rangle=G'(1)$. After implementing Line-operation, the line graph $\mathcal{G'}(\mathcal{V'},\mathcal{E'})$ has average degree equal to

\begin{figure}
\centering
\includegraphics[height=3.3cm]{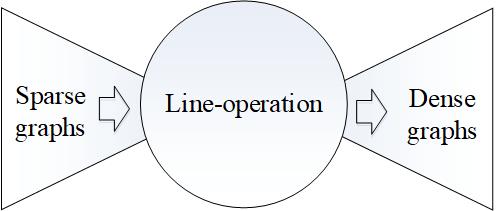}
\caption{\label{fig:epsart} The diagram of Line-operation for transforming a sparser graph into a denser one.}
\end{figure}

\begin{equation}\label{MF-3}
\langle k'\rangle=\frac{\langle k^{2}\rangle-\langle k\rangle}{\langle k\rangle}=\frac{G''(1)}{G'(1)}
\end{equation}
where $\langle k^{i}\rangle$ is the $i$-th moment of vertex degrees of graph $\mathcal{G}(\mathcal{V},\mathcal{E})$. Armed with the statements, certifying $\langle k'\rangle$ more than $\langle k\rangle$ is to show $G''(1)$ no less than $(G'(1))^{2}$. In most cases, the latter holds for a given graph, seeing Fig.1. Therefore, this provides us with an available manner in which we may densify the topological structure of a sparse graph as plotted in Fig.2.

On the other hand, the Line-operation can drastically damage many properties of initial graph $\mathcal{G}$ which are closely related to topological structure of the underlying graph, for instance, degree distribution and clustering coefficient. The topic of this paper is to construct dense graphs with power-law degree distribution. To this end, the most important is to choose an available graph $\mathcal{G}$ with an expected degree distribution that can be conveniently deduced to the power-law form using Line-operation in Fig.2. The selection of these such graphs will be successfully accomplished by adopting the other component of our novel framework.

From now on, let us divert insights into the development of the other component. First, for a given graph $\mathcal{G}(\mathcal{V},\mathcal{E})$, one can easily obtain its first-order subdivision graph $\mathcal{G}_{1}(\mathcal{V}_{1},\mathcal{E}_{1})$ by inserting one new vertex on each edge in $\mathcal{E}$, see Fig.3 for an illustrative example. Such an operation is called the \emph{first-order subdivision} in the jargon of graph theory. As we will demonstrate later, it is the first-order subdivision that guarantees the construction of seminal graphs that may be proven to satisfy those requirements mentioned above.

\begin{figure}
\centering
\includegraphics[height=6.5cm]{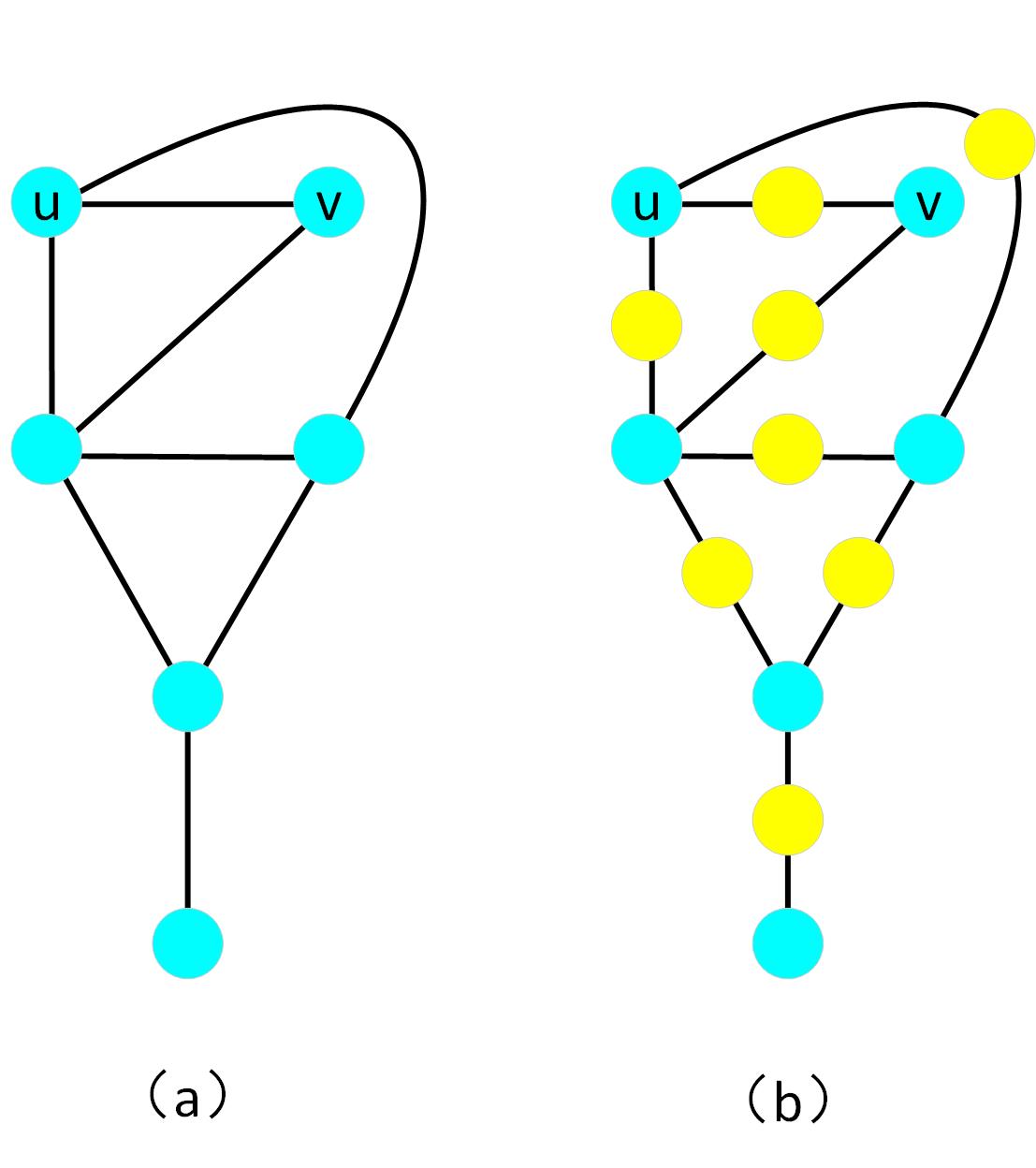}
\caption{\label{fig:epsart} The diagram of subdivision. Given an arbitrary graph $\mathcal{G}(\mathcal{V},\mathcal{E})$, if one inserts a new vertex $w$ to every edge $uv\in \mathcal{E}$ then the resulting graph, denoted by $\mathcal{G}_{1}(\mathcal{V}_{1},\mathcal{E}_{1})$, is called a \emph{first-order subdivision graph} of the original graph. To put this another way, such a graph $\mathcal{G}_{1}(\mathcal{V}_{1},\mathcal{E}_{1})$ can be obtained from graph $\mathcal{G}(\mathcal{V},\mathcal{E})$ by replacing every edge $uv\in \mathcal{E}$ by a unique path $uwv$ with length two.}
\end{figure}

By far, both components of our framework have been completely established. Now, our task is to clarify the concrete procedure of running our framework to obtain a dense graph with scale-free feature.

\textbf{\emph{Step 1}} For an arbitrary sparse graph $\mathcal{G}(\mathcal{V},\mathcal{E})$ whose degree distribution obeys
\begin{equation}\label{MF-4}
P(k)\sim k^{-\gamma}, \qquad 2<\gamma\leq3,
\end{equation}
if one applies the first-order subdivision to each edge in $\mathcal{E}$, then, the resulting graph is denoted by $\mathcal{G}_{1}(\mathcal{V}_{1},\mathcal{E}_{1})$.

\textbf{\emph{Step 2}} For the preceding graph $\mathcal{G}_{1}(\mathcal{V}_{1},\mathcal{E}_{1})$, one can manipulate Line-operation on each edge in $\mathcal{E}_{1}$. The end graph is regarded as graph $\mathcal{G'}_{1}(\mathcal{V'}_{1},\mathcal{E'}_{1})$.

Fig.4 illustrates the skeleton of our framework described here. Below provides a theoretical proof that shows that graph $\mathcal{G'}_{1}(\mathcal{V'}_{1},\mathcal{E'}_{1})$ is not only dense but also scale free. In addition, with the help of graph $\mathcal{G'}_{1}(\mathcal{V'}_{1},\mathcal{E'}_{1})$, some other properties of scientific interest will be discussed in the rest of this paper. Interestingly, some of which can be used as complementary materials to help us better understand the fundamental structural features of complex networks.

\section{Structural properties }

In this section, we aim at studying some main structural properties, for instance, degree distribution, of networked models $\mathcal{G'}_{1}(\mathcal{V'}_{1},\mathcal{E'}_{1})$ generated via the framework mentioned above.

\emph{Degree distribution}---Implementing the first-order subdivision will divide each edge $uv$ in $\mathcal{E}$ into two edges by inserting a new vertex $w$. As a result, the one of the both newborn edges above will connect old vertex $u$ with degree $k_{u}$ to degree $2$ vertex $w$ and the other connects degree $k_{v}$ vertex $v$ to young vertex $w$. To make further progress, the end graph $\mathcal{G'}_{1}(\mathcal{V'}_{1},\mathcal{E'}_{1})$ has degree distribution in form

\begin{equation}\label{MF-6}
P'(k)=kP(k)\sim k^{-\gamma'}, \qquad 1<\gamma'\leq2.
\end{equation}
This suggests that graph $\mathcal{G'}_{1}(\mathcal{V'}_{1},\mathcal{E'}_{1})$ obtained from scale-free graph $\mathcal{G}(\mathcal{V},\mathcal{E})$ by using our framework displays a power-law degree distribution. In other words, we indeed generate scale-free graphs with exponent falling into the interval from $1$ to $2$, which is what we wanted.

\begin{figure}
\centering
\includegraphics[height=3.5cm]{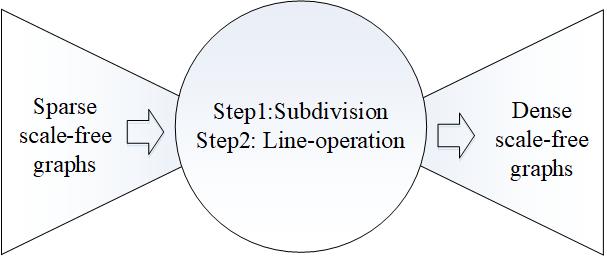}
\caption{\label{fig:epsart} The diagram of our framework for transforming a scale-free graph with sparsity structure into its corresponding dense graph with scale-free feature.}
\end{figure}

\emph{Average degree}---In the large graph size limit, it is straightforward using  Eqs.(\ref{MF-3}) and (\ref{MF-6}) to exactly calculate the solution for average degree $\langle k'\rangle$ in terms of

\begin{equation}\label{MF-7}
\langle k'\rangle\sim\int_{0}^{k'_{max}}kP'(k)dk\sim \left\{\begin{aligned}&\frac{1}{2-\gamma'}k'^{2-\gamma'}_{max}, \quad 1<\gamma'<2\\
&\zeta(1;k'_{max},k'_{min})\qquad \gamma'=2\end{aligned}
\right.
\end{equation}
where $k'_{min}$ and $k'_{max}$ represent, respectively, the expected minimum and largest degrees of vertices of graph $\mathcal{G'}_{1}(\mathcal{V'}_{1},\mathcal{E'}_{1})$, and symbol $\zeta(1;u,v)$ stands for Riemann zeta function with constraints, defined as $\zeta(1;u,v)=\sum_{i=v}^{u}i^{-1}$. It is worth noting that for finite number of vertices, the exact form of Eq.(\ref{MF-7}) can be found in Ref.\cite{C-I-D-G-2011}. As above, the expected largest degree value $k'_{max}$ is precisely equivalent to the greatest degree $k_{max}$ of vertices of graph $\mathcal{G}(\mathcal{V},\mathcal{E})$. In general, the expected value for $k_{max}$ can be asymptotically expressed with respect to the vertex number $|\mathcal{V}|$ as

 \begin{equation}\label{MF-8}
k_{max}\sim |\mathcal{V}|^{\frac{1}{\gamma-1}}.
\end{equation}

Plugging Eq.(\ref{MF-8}) into Eq.(\ref{MF-7}), for an arbitrary $\gamma'$ in range $(1,2]$, average degree $\langle k'\rangle$ will become infinite in the limit of large graph size. This means that scale-free graph $\mathcal{G'}_{1}(\mathcal{V'}_{1},\mathcal{E'}_{1})$ is by definition dense, which is what we want to see.

\emph{Diameter}---As the simplest of both measures for investigating small-world property of complex networks, diameter $D$ is the maximum among distances of all vertex pairs. As stated in \cite{R-C-2003}, scale-free networks are ultrasmall according to the relationship between diameter $D$ and vertex number $|\mathcal{V}|$. However, some networked models based on our framework will be able to exhibit the large diameter phenomenon as shown shortly. Now, if we suppose the diameter of an initial graph $\mathcal{G}(\mathcal{V},\mathcal{E})$ is equal to $D$, then the first-order subdivision will make the diameter $D_{1}$ of graph $\mathcal{G}_{1}(\mathcal{V}_{1},\mathcal{E}_{1})$ at most equivalent to two times $D$, i.e., $D_{1}\approx2D$. After that, it is clear to see that the diameter $D'$ of the end graph $\mathcal{G'}_{1}(\mathcal{V'}_{1},\mathcal{E'}_{1})$ will be approximately identical to diameter $D_{1}$ after applying Line-operation to graph $\mathcal{G}_{1}(\mathcal{V}_{1},\mathcal{E}_{1})$. In another words, the diameter $D'$ is magnitude of order the diameter $D$, namely,

 \begin{equation}\label{MF-9}
D'=O(D).
\end{equation}
This implies that if the seminal graph $\mathcal{G}(\mathcal{V},\mathcal{E})$ has small diameter, then the diameter of the resulting model $\mathcal{G'}_{1}(\mathcal{V'}_{1},\mathcal{E'}_{1})$  will be small. On the contrary, the large diameter of graph $\mathcal{G}(\mathcal{V},\mathcal{E})$ will ensure that the diameter of graph $\mathcal{G'}_{1}(\mathcal{V'}_{1},\mathcal{E'}_{1})$ is large. As reported in our prior work \cite{Ma-2019}, the growth scale-free networked model has a very large diameter ($D=2^{t}$, see \cite{Ma-2019} for a lot). Therefore, we can choose such a networked model as a seed and then obtain a dense graph with both scale-free feature and large diameter using our framework.

\emph{Clustering coefficient}---By definition, the clustering coefficient of a graph $\mathcal{G}$ can be written in the following form

 \begin{equation}\label{MF-10}
C=\frac{3\times triangle\, number}{number \,of \,connected\, triple},
\end{equation}
here a triangle is a cycle $C_{3}$ on three vertices and a connected triple means which a vertex is connected to a pair of other vertices. As described above, using the Line-operation, vertex $u$ with degree $k'_{u}$ of the resulting graph $\mathcal{G'}_{1}(\mathcal{V'}_{1},\mathcal{E'}_{1})$ will be allocated on a clique $K_{k'_{u}}$, a subgraph in which all vertex pairs are connected. Hence, the clustering coefficient $C'_{u}$ of vertex $u$ is calculated equal to $(k'_{u}-2)/k'_{u}$. According to Eq.(\ref{MF-6}), substituting the consequences calculated above into Eq.(\ref{MF-10}) produces

 \begin{equation}\label{MF-11}
C'=\frac{1}{|\mathcal{V}'|}\sum_{u\in \mathcal{V'}_{1}}C'_{u}\sim\int_{k'_{min}}^{k'_{max}}P'(k)C'_{k}dk\sim G(1).
\end{equation}
In the large graph size limit, $C'$ will approach the theoretical upper bound, i.e. unity.

Armed with Eqs.(\ref{MF-7}), (\ref{MF-9}) and (\ref{MF-11}), we can assert that when the seminal graph $\mathcal{G}(\mathcal{V},\mathcal{E})$ has scale-free feature and small-world property, the resulting graph $\mathcal{G'}_{1}(\mathcal{V'}_{1},\mathcal{E'}_{1})$ must be both scale free and small world. Furthermore, if the power-law exponent $\gamma$ of the initial graph falls in the range $(2,3]$, then the end graph has density structure.

\begin{figure*}
\centering
\subfigure[]{
\begin{minipage}[t]{0.45\linewidth}
\centering
\includegraphics[width=7cm]{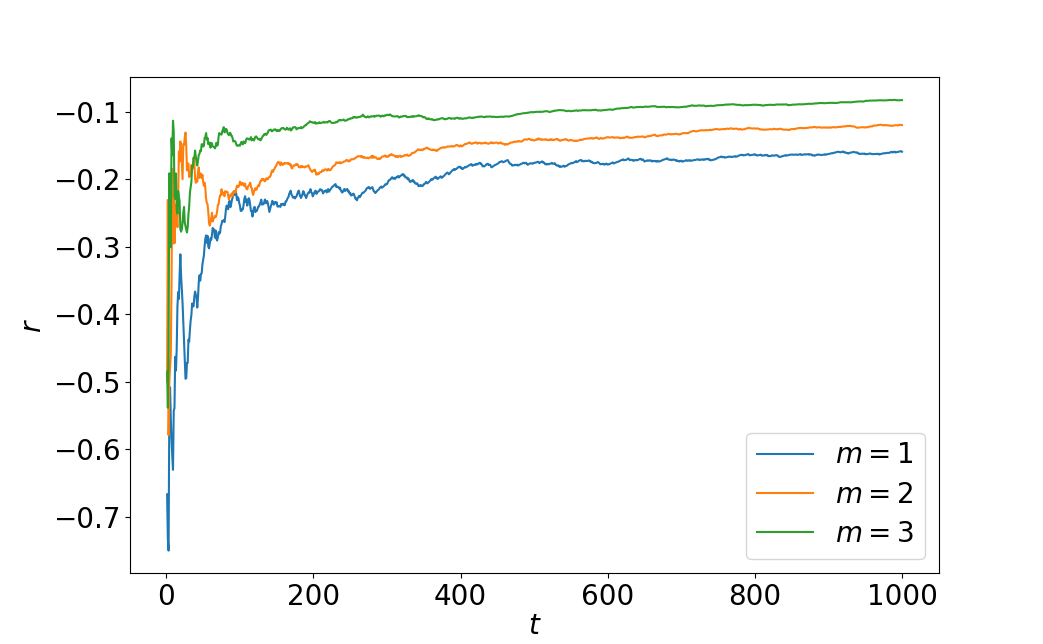}
\end{minipage}%
}%
\subfigure[]{
\begin{minipage}[t]{0.45\linewidth}
\centering
\includegraphics[width=7cm]{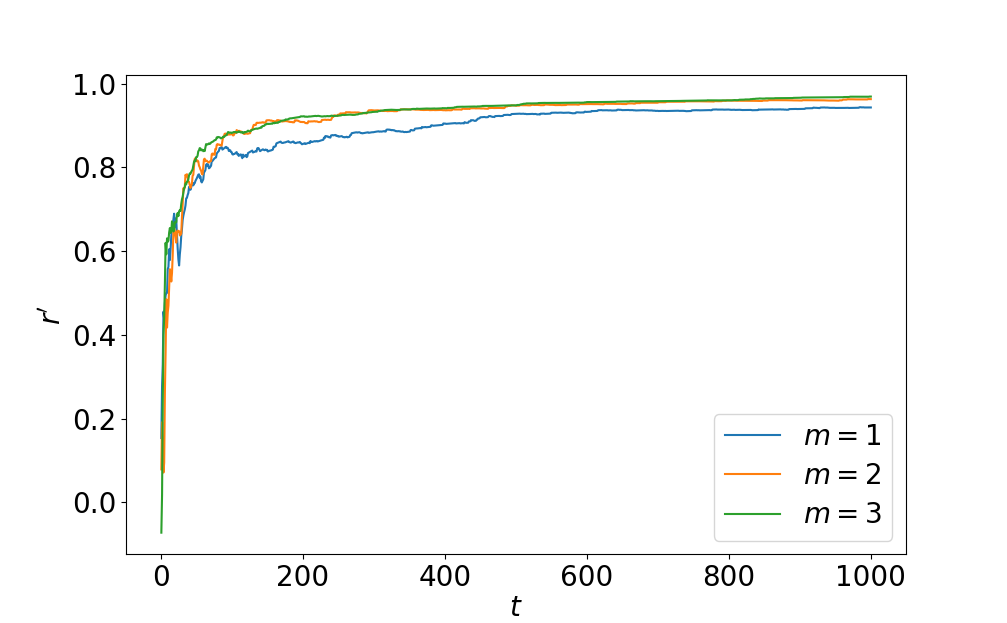}
\end{minipage}%
}%
\caption{\label{fig:wide} In the left panel, it is obvious to see that for different parameters $m$, which is number of connections to newly added vertex into the BA-model, Pearson correlation coefficient $r$ of the BA-model will always keep negative and approach $0$ in the limit of large graph size. On the contrary, in the right panel, Pearson correlation coefficients $r'$ of their corresponding graphs built by means of our framework are all positive and tend to the theoretical upper bound, i.e., unity.  }
\end{figure*}

\emph{Mixing structure}---Recent works have shown that in many networks, a number of vertices tend to be connected to other vertices like themselves. To analytically depict such a phenomenon, Newman in Ref \cite{Newman-2002} introduced a measure $r$, called Pearson correlation coefficient, which is defined in terms of

 \begin{equation}\label{MF-12}
r=\frac{|\mathcal{E}|^{-1}\sum\limits_{e_{ij}\in \mathcal{E}} k_{i}k_{j}-\left[|\mathcal{E}|^{-1}\sum\limits_{e_{ij}\in \mathcal{E}} \frac{1}{2}(k_{i}+k_{j})\right]^{2}}{|\mathcal{E}|^{-1}\sum\limits_{e_{ij}\in \mathcal{E}} \frac{1}{2}(k^{2}_{i}+k^{2}_{j})-\left[|\mathcal{E}|^{-1}\sum\limits_{e_{ij}\in \mathcal{E}} \frac{1}{2}(k_{i}+k_{j})\right]^{2}}.
\end{equation}
In practice, as stated in the process of calculating clustering coefficient, the resulting graph $\mathcal{G'}_{1}(\mathcal{V'}_{1},\mathcal{E'}_{1})$ contains various types of cliques as subgraphs. As an immediate consequence, the corresponding $r'$ will be greater than the $r$ of initial graph $\mathcal{G}(\mathcal{V},\mathcal{E})$. In order to make our statements more concrete, we make use of the well-known scale-free graph, the BA-model due to Barabasi \emph{et al} in \cite{Albert-1999-1}, as a seed to generate its corresponding dense graphs. In principle, such simulations should use a good unbiased sampling algorithm as reported in \cite{Genio-2010}. In essence, the simulations above may be adequately adopted to illustrate that the theoretical analysis introduced herein is sound. As with the BA-model, we obtain solutions for Pearson correlation coefficients for six scale-free graphs by varying the number $m$ of edges originating from each newly added vertex, see Fig.5 for more details. It is worth noticing that as demonstrated in recent literature \cite{Johnson-2010} and \cite{Williams-2014}, almost all scale-free networks have been proven to show disassortativity from various respects including entropy. Hence, our work is to build up a way to produce such rare types of scale-free networks without disassortativity.

\emph{Community structure}---As the final topological measure discussed in our work, community structure,
within which there is a higher density of edges and between which there is a few, has been a focus of current
researches of significant interest, particularly within statistical physics and computer science \cite{Fortunato-2010}. Here, we utilize one approach in widest current use, i.e, modularity maximization, to investigate the community structure of the resulting graph $\mathcal{G'}_{1}(\mathcal{V'}_{1},\mathcal{E'}_{1})$ and further probe the distribution of community size $s'$ on graph $\mathcal{G'}_{1}(\mathcal{V'}_{1},\mathcal{E'}_{1})$.

The modularity of a graph $\mathcal{G}(\mathcal{V},\mathcal{E})$ is given by

 \begin{equation}\label{MF-13}
Q=\frac{1}{2|\mathcal{E}|}\sum_{ij}\left[A_{ij}-\frac{k_{i}k_{j}}{2|\mathcal{E}|}\right]\delta_{g_{i},g_{j}}
\end{equation}
where $k_{i}$ is the degree of vertex $i$, $\delta_{i,j}$ is the Kronecker delta as above, $g_{i}$ represents the community to that vertex $i$ belongs, and $A_{ij}$ is the element of the adjacency matrix of graph $\mathcal{G}(\mathcal{V},\mathcal{E})$ which is equal to $1$ when vertex $i$ is adjacent to vertex $j$ and $0$ otherwise. Before processing the following calculations, let us recall the construction of $\mathcal{G'}_{1}(\mathcal{V'}_{1},\mathcal{E'}_{1})$ and can evidently see that the sizes of various types of cliques is in spirit similar to the degree sequence of initial $\mathcal{G}(\mathcal{V},\mathcal{E})$. With the help of two assertions in \cite{UB-2009}, (\emph{Assertion 1} In a maximum modularity clustering of graph $\mathcal{G}(\mathcal{V},\mathcal{E})$, none of the cliques $H_{1}; ... ; H_{k}$ is split. \emph{Assertion 2} In a maximum modularity clustering of $\mathcal{G}(\mathcal{V},\mathcal{E})$, every cluster contains at most one of the cliques $H_{1}; ... ; H_{k}$.) we can confirm that the community size $s'$ distribution has a power-law form

 \begin{equation}\label{MF-14}
P(s')\sim s'^{\gamma}, \qquad 2<\gamma\leq3.
\end{equation}
Surprisingly, such a phenomenon has been discovered in some real-world complex networks, such as, Amazon copurchasing network in \cite{Clauset-2004}.

\section{Discussion and conclusion }

\emph{Discussion}---We have introduced a novel framework for producing a dense graph with scale-free feature from a given sparse graph whose degree distribution obeys the power-law form. From the theoretical point of view, the resulting graphs based on our framework can be selected to provide helpful materials for understanding the construction of scale-free graphs with density feature, a class of networks that are rarely observed in the study of complex networks. In addition, these resulting graphs also display some other interesting topological properties unseen in most scale-free graphs, such as, the higher clustering efficient shown in Eq.(\ref{MF-11}) and the stronger assortative structure plotted in Fig.5. Last but not the least, the community size distributions of the resulting graphs may share an identical form with the degree distribution of the corresponding graphs, as revealed in Eq.(\ref{MF-14}).

In conclusion, our theory can provide another class of generative method for establishing some desirable graphs of scientific interest. And then, these findings can lead us to test meaningful hypotheses in an evolving networked model, particularly, in the scale-free graph.

\section*{\label{sec:level1}ACKNOWLEDGMENTS}
The authors would like to thank Bing Yao for useful conversations. The research was supported in part by the
National Key Research and Development Plan under grant 2017YFB1200704 and the National Natural
Science Foundation of China under grant No. 61662066

\end{document}